\documentstyle[12pt]{article}

\title{\textbf{Conjecture on the Avoidance of the Big Crunch}}
\author{\textbf{Cheng-Yi Sun\footnote{cysun@mails.gscas.ac.cn}
\ and \ De-Hai Zhang\footnote{dhzhang@gscas.ac.cn}
}\\
Department of Physics,\\
The Graduate School of The Chinese Academy of Sciences,\\
Beijing 100049, P.R.China.}
\input{tcilatex}

\begin{document}

\maketitle
\begin{abstract}
In \cite{KKLT}, KKLT give a mechanism to generate de Sitter vacua
in string theory. And recently, the scenario, {\em landscape}, is
suggested to explain the problem of the cosmological constant. In
this scenario, the cosmological constant is a de Sitter vacuum.
The vacuum is metastable and would decay into an anti-de Sitter
vacuum finally. Then the catastrophe of the big crunch appears. In
this paper by conjecturing the physics at the Planck scale, we
modify the definition of the Hawking temperature. Hinted by this
modification, we modify the Friedmann equation. we find that this
avoid the singularity and gives a bouncing cosmological model.
\end{abstract}



\section{Introduction}

The data from the observation of the first year Wilkinson
Microwave Antisotropy Probe (WMAP) \cite{a0302207,a0302209} and
the observation of the SNe Ia \cite{Ia SuperNova} make us almost
be sure that the expansion of the universe is accelerating. The
simplest explanation is that a small but non-zero cosmological
constant, the de Sitter vacuum, dominates the present universe.
Recent years, great efforts basing on the string theory have been
paid to solve the cosmological constant problem and to construct a
complete process of the cosmological evolution. In\cite{KKLT}, a
mechanism, KKLT mechanism, is given to get de Sitter vacua in
string theory. And a scenario named {\em Landscape} has been
suggested \cite{t0302219,t0501082}. In this scenario it is argued
that string theory has a landscape of vacua. The supersymmetric
(SUSY) sector of the landscape has the zero vacuum energy. The
non-SUSY sector has the stochastic distribution of vacua energies
around the zero vacuum energy, where some vacua are de Sitter
vacua with positive vacuum energy and others are anti-de Sitter
vacua with negative vacuum energy. One of the de Sitter vacua
describes the present acceleration of our unverse, and this vacuum
is a metastable state. It would decay into another vacuum with a
lower cosmological constant. Unfortunately, the detailed
information of the landscape is absent. So the exact process of
the decay is unknown. However, due to the exist of the anti-de
Sitter vacua, we are sure that the final vacuum should be an
anti-de Sitter vacuum. On the other hand, it is familiar to us
that, in cosmology, the decay of dS vacua to AdS vacua would cause
the big crunch inside the vacuum-decay bubble.

The catastrophe of the big crunch seems to be inevitable and
discouraging. Even, this is one of the reasons that lead Banks to
conjecture that the dS vacua are stable in sensible gravity
theories \cite{t0211160}. However, this may be solved in another
path. We know that the energy density of the universe increases as
the contraction. It must have reached the Planck energy scale
before the singularity is reached. On the other hand, the physical
laws which is used to derive the appearance of the singularity is
applicable only to the energy scale much lower than the Planck
energy scale. So we think the singularity of the big crunch may be
eliminated due to the new physics near the Planck scale.
Unfortunately, the physics near the Planck scale is unclear up to
now. But this can not block us to conjecture the physics at Planck
scale.

In this paper, we first investigate the Hawking radiation of a
black hole. We find that the concept of the Hawking temperature is
questionable for a black hole with the Planck mass. By modifying
the definition of the Hawking temperature we solve it. Then,
analogously, we modify the Friedmann equation. This modification
avoid the singularity of the big crunch by giving a bounce at the
Planck scale.

\section{the Hawking temperature of black holes at the Planck
scale}

In this section we investigate the Hawking radiation of a black
hole. First, we consider a Schwarzschild black hole with mass $m$.
Hereafter, we take $8\pi G=M_{pl}^{-2}$. Then the radius of the
horizon of the black hole is $r_H=\frac{m}{4\pi M_{pl}^2}$ and the
mass density of the black is $\rho= \frac{3}{4\pi}mr_H^{-3}$.
Hawking has shown (1974, 1975) that a distant observer will detect
a thermal spectrum of particles coming from the black hole, at a
temperature $T=M_{pl}^2m^{-1}$. This is just the well-known
Hawking radiation. When the mass is at the scale of a star, the
temperature is very low and the concept of the temperature makes
sense. However, we know the mass of the black hole decreases as
the radiation processing. When the mass approaches the Planck
scale, the temperature approaches the Planck scale, too. This
means the average energy of the radiation particles, $\sim T$,
would be about the Planck energy, which is nearly equal to the
mass of the black hole. It implies that one radiation particle
carries away nearly the whole energy of the black hole. What does
this mean? In fact this is just the translation of the black hole
and no radiation exists at this time. So we expect that the
radiation temperature should disappear at the Planck scale. But
the semiclassical definition of the Hawking temperature give a
very high temperature. The reason ,we think, is the backreaction
effects of the emitted particles are neglected in the definition.
However, from the analysis above, we know that the backreaction is
not negligible when the mass of a black hole approaches the Planck
scale.

So let's consider the backreaction effects. We know the nonzero
Hawking temperature produce the Hawking radiation, with the
radiation energy density $\rho_R\sim T^4$. According to the three
equations given in the last paragraph, we know $T^2\sim
M_{pl}^{-2}\rho$. So we get $\rho_R\sim M_{pl}^{-4}\rho^2$.
Naturally, this radiation would reduce the energy density of the
black hole. Then considering this effect, we expect the
temperature should be $T^2\sim
M_{pl}^{-2}(\rho-\rho_R)=M_{pl}^{-2}(\rho-M_{pl}^{-4}\rho^2)$.
Now, as a toy model, we conjecture naively that the temperature of
the black hole should be modified as
\begin{equation}
\label{temperature}T^2=\frac{\rho}{48\pi^2
M_{pl}^2}(1-\frac{\rho}{M_{pl}^4})
\end{equation}
When the mass is much larger than the Planck mass, the density is
much smaller than the Planck scale. Then the second term in the
bracket on the right-hand side is negligible and the modified
equation is equivalent to the ordinary definition approximately.
But the difference is remarkable as the mass approaches the Planck
scale. It is obvious that Eq.(\ref{temperature}) gives $T=0$ at
$\rho=M_{pl}^4$. So the difficulty in the last paragraph does not
exist. In addition, this expression implies the temperature has
the maximum, $T_m\sim \frac 14M_{pl}$. This is consistent with the
concept of the Hagedorn temperature in string theory.

\section{the evolution of the universe at the Planck scale}

From the investigation in the black hole above, we know the
character of the physics changes remarkably near the Planck scale.
We know, a de Sitter universe with a cosmological constant,
$\Lambda$, is similar to a black hole. It has also a temperature,
$T\sim H$. The Hubble parameter, $H$, is governed by the Friedmann
equation
\begin{equation}
\label{dSFriedmannEq}H^2=\frac{\dot
a}{a}=\frac{1}{3M_{pl}^2}\Lambda,
\end{equation}
where $a$ denotes the scale factor in the FRW metric and ''dot''
denotes derivative with respect to the physical time. Hinted by
Eq.(\ref{temperature}), we conjecture that
Eq.(\ref{dSFriedmannEq}) should also be modified as
\begin{equation}
\label{ModidSFriedmannEq}H^2=\frac{1}{3M_{pl}^2}\Lambda(1-\frac{\Lambda}{M_{pl}^4}).
\end{equation}
Extending this modification to a general FRW universe directly, we
get the modified Friedmann equation
\begin{equation}
\label{ModFriedmannEq}H^{2}=\frac{1}{3M_{pl}^2}\rho(1-\frac{\rho}{M_{pl}^4}),
\end{equation}
Obviously the second term in the bracket on the right-hand side is
negligible when the energy density is much lower than the Planck
scale. Then this modified Friedmann equation becomes the standard
Friedmann equation. But, at the hight energy density, the second
term play a key role. It is obvious that this modified Friedmann
equation avoid the singularity of the big crunch by giving a
bounce of the universe at $\rho=M_{pl}^4$.

Now let's analyze the evolution of the universe in detail as
$\rho$ approaching $M_{pl}^4$ in the big crunch. We assume the
equation of state of the dominant component of the universe is
$p=w\rho$. Here, we emphasize that we only consider the component
with $w>1$. We define the scale factor $a(t)=a(t_p)+x$, where $x$
is an infinitesimal parameter and $t_p$ denotes the moment of the
bounce, $\rho=M_{pl}^4$. Then, as $\rho$ approaching $M_{pl}$, we
get
\begin{equation}
\rho=M_{pl}^4[1-\frac{3(1+w)x}{a(t_p)}], \ \ H^2=\frac{\dot
x^2}{a(t_p)^2}
\end{equation}
Substituting these into Eq.(\ref{ModFriedmannEq}), we get
\begin{equation}
\frac{dx}{dt}=-\sqrt{M_{pl}^2(1+w)a(t_p)x}.
\end{equation}
Solving this equation, we get
\begin{equation}
x=\frac{M_{pl}^2(1+w)a(t_p)}{4}(t-t_p)^2.
\end{equation}
Now we are sure that the bounce at $t_p$ is smooth according to
Eq.(\ref{ModFriedmannEq}). So the catastrophe of the big crunch is
eliminated. After the bounce, the universe begins to expand again.
Naturally, we may take the bounce as a big bang after which the
new evolution of the universe begins.

The similar modification has been shown in another view,
braneworld scenarios \cite{g0208047}. The difference is that in
braneworld cosmology this is taken as the effect of the
extradimensions. Here we suppose that this is due to the effect of
the physics near the Planck scale.

\section{Summary}

In this paper, in order to solve the disaster of the big crunch,
we conjecture the physics at the Planck scale. We first
investigate the concept of the Hawking temperature of the black
hole when the mass of the black hole decreases to the Planck mass.
We find some paradox appears. We try to solve it by modifying the
definition of the temperature. Analogously, by modifying the
Friedmann equation, we avoid the singularity of the big crunch. Of
course, our modification is just a toy model. We do not think we
have obtained the law of physics at the Planck scale. We just take
it as an interesting attempt. Finally, we emphasize that many
concepts at lower energy scale can not be used directly to
describe the physics at the Planck scale.

\end{document}